\documentclass[a4paper]{article}

\usepackage{amsmath, amssymb, amsfonts, amsthm}
\usepackage{epsfig}
\usepackage{graphicx}  % standard LaTeX graphics tool
\usepackage{color}     % for color figures
\usepackage{lscape}
\usepackage{color}
\begin{document}

\title{\Large \bf Modified Chaplygin Gas Cosmology with Bulk Viscosity}
{\small
\author{H. B. Benaoum \\
 Department of Applied Physics, University of Sharjah, \\
P.O. Box 27272, Sharjah, United Arab Emirates \\
Email: hbenaoum@sharjah.ac.ae}
}

\maketitle
\begin{center}
\small{\bf Abstract}\\[3mm]
\end{center}
We investigate the viscous modified Chaplygin gas cosmological model. Solutions for different values of the viscosity parameter are obtained using both analytical and numerical methods. 
We have calculated the deceleration and defined {\em newly} statefinder $\{r, s \}$ pair in D dimensions. 
It is shown that when $D=4$, the usual statefinder parameters are recovered. Furthermore, we apply the statefinder diagnostic to the MCG model with and without viscosity in $D$ dimensions and explore these parameters graphically.
\\\\\

%{\bf PACS numbers}: 98.80.-k, 98.80.Cq,98.80.Jk, 95.36.+x
\begin{minipage}[h]{14.0cm}
\end{minipage}
\vskip 0.3cm \hrule \vskip 0.5cm
%%%%%%%%%%%%%%%%%%%%%%%%%%%%%%%%%%

\newpage
\section{Introduction} 
A growing number of observational data indicate that the observable universe is undergoing a phase of accelerated expansion \cite{per1}-\cite{spergel}. 
The source of this cosmic acceleration is attributed to an unknown dark energy component with a negative pressure, which dominates the universe at recent cosmological time. \\

Several mechanisms have been proposed to describe the physical nature of this dark energy component; among them, the single-component fluid known as Chaplygin 
gas has attracted a lot of interest in recent times \cite{peebles1}-\cite{dvali}. 
Many variants of the  Chaplygin gas  model have been proposed in the literature. 
A further general model named modified Chaplygin gas (MCG) has been introduced by Benaoum \cite{benaoum} and obeys the following equation of state : 
\begin{eqnarray}
p & = & A \rho - \frac{B}{\rho^{\alpha}}  
\end{eqnarray}
where  $A, B$ and $\alpha$ are universal positive constants. \\ 
The equation of state with $A = 0$ and $\alpha = 1$ known as Chaplygin gas (CG) was first introduced by Chaplygin to study the lifting forces on 
a plane wing in aerodynamics. Its generalized form with $A = 0$ and $\alpha > 0$ is known as the generalized Chaplygin gas (GCG) was first introduced by 
Kamenshchik {\em et al.} \cite{kamenshchik} and Bento {\em et al.} \cite{bento} . \\

The MCG has the remarkable property of describing the dark sector of the universe (i.e. dark energy and dark matter) as a single component that acts 
as both dark energy and dark matter. It interpolates from matter-dominated era to a cosmological constant-dominated era. 
It is well known that the CG, MCG and theirs variants have bee extensively studied in the literature  \cite{debnath}-\cite{benaoum1}. 
Several papers discussing various aspects of the behavior of MCG to reconcile the standard cosmological model with observations have been considered \cite{lu1}-\cite{fabris}. \\

On the other hand, dissipative effects, including both bulk and shear viscosity, play an important role in the evolution of the universe.
In viscous cosmology, shear viscosities arise in relation to space anisotropy while the bulk viscosity accounts for the space isotropy. 
In the study of cosmology, shear viscosities are ignored since the cosmic microwave background (CMB) does not indicate significant anisotropies, and only bulk viscosities are taken into account for the viscous fluid. 
Furthermore, bulk viscosity related to a grand unified theory phase transition may lead to an explanation of the accelerated cosmic expansion. 
Bulk viscous cosmological models have been discussed by several authors \cite{barrow}-\cite{singh2}. Chaplygin gas with viscosity has been considered in \cite{zhai}. Very recently, FRW MCG viscous fluid cosmological model has been studied \cite{saadat}. \\

In order to differentiate between various dark energy models, a geometrical diagnostic, called statefinder $\{r,s\}$ pair, has been introduced in \cite{sahni}. 
It involves the third derivative of the scale factor $a$ with respect to the time, just like the Hubble and 
deceleration parameters involve the first and second derivative respectively. 
The statefinder pair defines two new cosmological parameters $r$ and $s$ 
in addition to the Hubble parameter $H$ and the deceleration parameter $q$ as follows :
\begin{eqnarray}
r & = & \frac{\dddot{a}}{a H^3} \nonumber \\
s & = & \frac{r-1}{3 (q - \frac{1}{2})} ~~~~.
\end{eqnarray} 
For the $\Lambda$CDM model, the statefinder diagnostic pair $\{r,s \}$ takes the fixed point $\{1,0 \}$ whereas for the standard cold dark matter (SCDM), it corresponds to the value $\{1,1 \}$ \cite{sahni}-\cite{alam}. Departure 
of a dark energy model from the $\Lambda$CDM fixed point is a good way of establishing the {\em distance} of 
this model from the flat $\Lambda$CDM. \\

This work is organized as follows. 
The physical model of the viscous cosmological MCG and the basic equations are presented in section 2.  
Exact solutions for viscosity parameter $\nu \neq \frac{1}{2}$ and $\nu = \frac{1}{2}$ are obtained using both analytical and numerical methods.  
In section 3, the deceleration parameter and the statefinder parameters in D dimensions are calculated.It is shown that when $D=4$, the usual statefinder parameters are recovered. 
Based on this, we apply the statefinder diagnostic to the viscous MCG model in $D$ dimensions. The statefinder parameters $\{r,s \}$ are derived and the 
numerical results are presented.
Finally, this work ends with discussion and concluding remarks in section 4. 

%%%%%%%%%%%%%%%%%%%%%%%%%%%%%%%%%%%%%%%%%%%%%%%%%%%%%%%%%%%%%%%%%%
\section{Viscous Modified Chaplygin Gas}
We consider a Friedmann Robertson-Walker (FRW) universe described by the following metric : 
\begin{eqnarray}
ds^2 & = & - dt^2 + a^2 (t) \left[ \frac{dr^2}{1 - k r^2} +r^2 d \Omega^2_{D-1} \right] ~~~~~.
\end{eqnarray}
Here $a (t)$ is the scale factor of the universe and the curvature $k = 0, \pm 1$ describes spatially flat, closed or open spacetimes respectively. \\

The Einstein's field equation is given by : 
\begin{eqnarray}
R_{\mu \nu} - \frac{1}{2} g_{\mu \nu} R & = & T_{\mu \nu} 
\end{eqnarray}
where $T_{\mu \nu}$ is the energy-momentum tensor. \\

We assume that the spacetime is filled with only one component fluid having a bulk viscosity. In this case, the energy-momentum tensor can be written 
as follows : 
\begin{eqnarray}
T_{\mu \nu} & = & p_{eff} ~ g_{\mu \nu} + \left( \rho + p_{eff} \right)~ u_{\mu} u_{\nu} 
\end{eqnarray}
where $u_{\mu}$ is the velocity, and the effective pressure $p_{eff}$ can be expressed as follows : 
\begin{eqnarray}
p_{eff} & = & p + \Pi 
\end{eqnarray}
which is the sum of the equilibrium pressure $p$ and the bulk pressure $\Pi$. The bulk viscous pressure $\Pi$ is represented by the Eckart's expression 
which is proportional to the Hubble parameter $H$ with proportionality factor identified as the bulk viscosity coefficient $\xi$ : 
\begin{eqnarray}
\Pi & = & - (D-1) H \xi ~~~~~. 
\end{eqnarray}
In most of the investigations in cosmology, the bulk viscous coefficient is assumed to have a power law dependence on the energy density, 
\begin{eqnarray}
\xi & = & \xi_0 ~\rho^{\nu} 
\end{eqnarray}
where $\xi_0 >0$ and $\nu$ are constants. \\

The Friedmann equations which govern the evolution of the scale factor are given by : 
\begin{eqnarray}
H^2 & = & \frac{2 ~\rho}{(D-1) (D-2)} - \frac{k}{a^2} 
\end{eqnarray}
\begin{eqnarray}
\dot{H} & = & \frac{1}{D-2} ( \rho + p_{eff} ) + \frac{k}{a^2}
\end{eqnarray}
and the conservation law equation reads :
\begin{eqnarray} 
\dot{\rho} + (D-1) ~H ( \rho + p_{eff} ) & = & 0 
\end{eqnarray}   
with $H = \frac{\dot{a}}{a}$ being the Hubble parameter. Here,we assume that $c = 1$ and $8 \pi G = 1$.  \\

Using the equation of state of the MCG with bulk viscosity in the above conservation equation, we get : 
\begin{eqnarray}
\dot{\rho} + (D-1) H \left( (A+1) \rho - \frac{B}{\rho^{\alpha}} - (D-1) H \xi_0 \rho^{\nu} \right) & = & 0 ~~~~. 
\end{eqnarray}
In the following subsections, this equation is studied for different values of the viscosity parameter $\nu$. In particular, we look for 
cosmological solutions corresponding to the viscosity parameter $\nu \neq \frac{1}{2}$ and $\nu = \frac{1}{2}$. 

\subsection{Solution with $\nu \neq \frac{1}{2}$}
Here, exact solution for $\nu \neq \frac{1}{2}$ is studied using both analytical and numerical methods. The evolution equation 
(i.e. equation (12)) for the energy density $\rho$ for the viscous dissipative MCG flat homogeneous cosmological model can be written as :
\begin{eqnarray}
\frac{\rho^{\alpha} d \rho}{(A+1) \rho^{\alpha+1} - B - c_D \xi_0  ~\rho^{\alpha + \nu +\frac{1}{2}}} + \frac{d a^{D-1}}{a^{D-1}} & = & 0 ~~
\end{eqnarray} 
where we denote $c_D = \sqrt{\frac{2 (D-1)}{D-2}}$. \\

Because of the complicated nonlinear character of the differential equation (13), we expand it in power series of $\xi_0$. Equation (13) becomes : 
\begin{eqnarray}
\rho^{\alpha} d \rho ~\sum_{n=0}^{\infty} \frac{c_D^n \xi_0^n \rho^{n (\alpha +\nu + \frac{1}{2})}}{\left( (A+1) \rho^{\alpha+1} -B \right)^{n+1}} +  
\frac{d a^{D-1}}{a^{D-1}} & = & 0 ~~~~~.
\end{eqnarray}

By performing the change of variable $u = \rho^{\alpha+1}$, the latter equation takes the form : 
\begin{eqnarray}
\frac{1 }{\alpha +1} ~\sum_{n=0}^{\infty} ~\frac{(-1)^{n+1} c_D^n \xi_0^n}{B^{n+1}} ~
\frac{ d u^{1+ n \frac{\alpha+\nu+\frac{1}{2}}{\alpha+1}}}{\left(1-  \frac{A+1}{B} u \right)^{n+1}} +  \frac{d a^{D-1}}{a^{D-1}} & = & 0 ~~~~~.
\end{eqnarray}
The general solution of equation (15) is given in terms of the hypergeometric function ${}_2F_1$. After integration, we get : 
\begin{eqnarray}
\frac{\ln ( (A+1) u -B )}{(\alpha+1)(A+1)} + \ln a^{D-1}  + \frac{1}{\alpha +1} ~\sum_{n=1}^{\infty} ~\frac{(-1)^{n+1} c_D^n \xi_0^n}{B^{n+1}} ~
\frac{ u^{1+ n \frac{\alpha+\nu+\frac{1}{2}}{\alpha+1}}}{1+n \frac{\alpha +\nu+\frac{1}{2}}{\alpha+1}} ~
{}_2F_1 \!\left[ \begin{array}{cc}
1+n ,1+ n \frac{\alpha+\nu+\frac{1}{2}}{\alpha + 1} \\
~~~~~~~~2+ n \frac{\alpha+\nu+\frac{1}{2}}{\alpha + 1} \end{array}  ;
\frac{A+1}{B} u \right] ~=~ C_0
\end{eqnarray}
where $C_0$ is an integration constant. \\

Now, by using the nth-derivative formula of the hypergeometric function ${}_2F_1$ which is given by : 
\begin{eqnarray}
\frac{d^n}{d z^n} {}_2F_1 \!\left[ \begin{array}{cc}
a ,b \\
~~~c \end{array}  ;
z \right] & = & \frac{(a)_n (b)_n}{(c)_n} {}_2F_1 \!\left[ \begin{array}{cc}
a+n ,b+n \\
~~~~~~~~c+n \end{array}  ;
z \right]
\end{eqnarray}
we further rewrite the equation (16) as :
\begin{eqnarray}
\frac{\ln ( (A+1) u -B )}{(\alpha+1)(A+1)} + \ln a^{D-1} + \frac{1}{\alpha +1} ~\sum_{n=1}^{\infty} ~\frac{(-1)^{n+1} c_D^n \xi_0^n}{B (A+1)^{\alpha+1}} ~
\frac{ u^{1+ n \frac{\alpha+\nu+\frac{1}{2}}{\alpha+1}}}{n! (1+n \frac{\nu-\frac{1}{2}}{\alpha+1})} ~
\frac{d^n}{d u^n} {}_2F_1 \!\left[ \begin{array}{cc}
1 ,1+ n \frac{\nu-\frac{1}{2}}{\alpha + 1} \\
~~~2+ n \frac{\nu-\frac{1}{2}}{\alpha + 1} \end{array}  ;
\frac{A+1}{B} u \right] ~=~ C_0 ~~.
\end{eqnarray}

From the above equation, the density $\rho$ can be easily expresses as : 
\begin{eqnarray}
\rho^{\alpha+1} & = & \frac{B}{A+1} + \frac{C ~\prod\limits_{n=1}^{\infty} e^{\left(-\frac{c_D \xi_0}{A+1}\right)^n ~\frac{f_n (\rho)}{n!}}}{a^{(D-1) (\alpha+1) (A+1)}}
\end{eqnarray} 
where $C= e^{C_0}$ is the integration constant and we define the function $f_n (\rho)$ as : 
\begin{eqnarray}
f_n (\rho) & = & \frac{A+1}{B} \frac{\rho^{\alpha +1 + n (\alpha+\nu+\frac{1}{2})}}{1 + n \frac{\nu-\frac{1}{2}}{\alpha +1}} ~ 
\left(\frac{d}{d \rho^{\alpha+1}} \right)^n {}_2F_1 \!\left[ \begin{array}{cc}
1 ,1+ n \frac{\nu-\frac{1}{2}}{\alpha + 1} \\
~~~2+ n \frac{\nu-\frac{1}{2}}{\alpha + 1} \end{array}  ;
\frac{A+1}{B} \rho^{\alpha+1} \right] ~~~~.
\end{eqnarray} 

It is easy to see that when $\xi_0 = 0$ in equation (20), the usual MCG is recovered. In Fig. 1, we plot the 
evolution of the energy density $\rho$ with respect to the scale factor $a$ where the viscosity parameter $\xi_0$ 
varies. We have considered three values $\xi_0 = 0$ (black color), $0.01$ (red color) $0.03$ (blue color), respectively where we set the parameters to be $\alpha = 1, A=\frac{1}{3},B =1,\nu = \frac{3}{2}$ and $D=4$. 
Fig. 1 shows that as $\xi_0$ increases the density $\rho$ decreases. \\

\begin{figure}[hbtp]
\centering
\epsfxsize=6cm
\centerline{\epsfbox{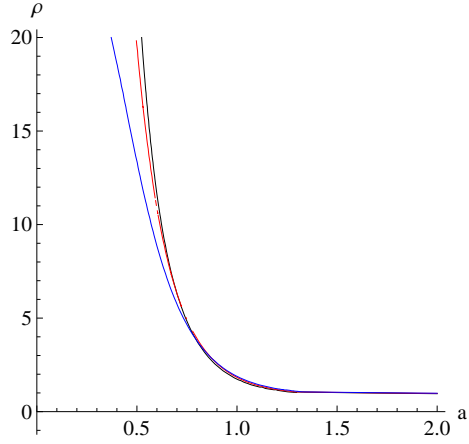}}
\caption{Plot of the density $\rho$ versus the scale factor $a$ for $\xi_0 = 0, 0.01, 0.03$.}
\end{figure} 
In Fig. 2, the variation of the energy density is plotted by fixing the value of the viscosity parameter $\xi_0 = 0.01$ and varying the parameter 
$\nu = 0$ (black color) and $3/2$ (red color). \\

\begin{figure}[hbtp]
\centering
\epsfxsize=6cm
\centerline{\epsfbox{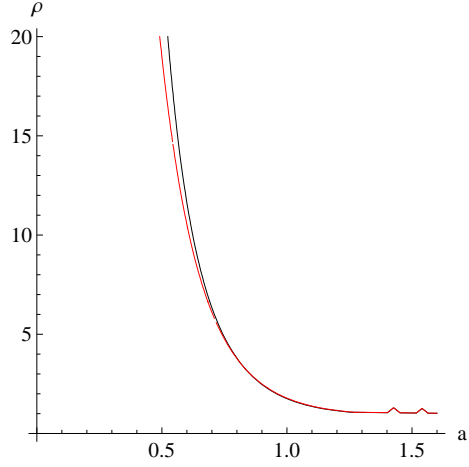}}
\caption{Plot of the density $\rho$ versus the scale factor $a$ for $\nu = 0, 3/2$.}
\end{figure} 

From equation (12), we get the explicit form of the time $t$ in terms of the energy density $\rho$ as : 
\begin{eqnarray}
c_D ~t + C  & = & \sum_{n=1}^{\infty} \left(- \frac{c_D \xi_0}{A+1} \right)^n 
\frac{ \rho^{\alpha-\frac{1}{2}+ n (\alpha+\nu+\frac{1}{2})}}{B~ n! \left(1- \frac{1}{2 (\alpha+1)}+ n \frac{\nu-\frac{1}{2}}{\alpha+1} \right)} 
\left(\frac{d}{d \rho^{\alpha+1}} \right)^n {}_2F_1 \!\left[ \begin{array}{cc}
1 ,\beta_n \\
~~~~~~~1+\beta_n \end{array}  ;
\frac{A+1}{B} \rho^{\alpha+1} \right] 
\end{eqnarray}
where $C$ is an integration constant and we denote $\beta_n = 1- \frac{1}{2 (\alpha+1)}+ n \frac{\nu-\frac{1}{2}}{\alpha + 1}$. 

\newpage
\subsection{Solution with $\nu = \frac{1}{2}$}
By rescaling the density as 
$\bar{\rho} = \rho ~a^{(D-1) (A+1 - c_D~\xi_0)}$, the equation (12) for flat case becomes :
\begin{eqnarray}
\bar{\rho}^{\alpha} ~\frac{d \bar{\rho}}{d a^{D-1}} & = & B ~a^{(D-1) (\alpha +1) (A + 1 - c_D \xi_0)}  ~~~~.
\end{eqnarray}
Integrating the latter equation leads to :
\begin{eqnarray}
\frac{\bar{\rho}^{\alpha +1}}{\alpha +1} & = & \frac{B ~a^{(D-1) (\alpha +1) (A + 1 - c_D \xi_0)}}{(\alpha +1) (A + 1 - c_D\xi_0)} + \frac{C}{\alpha +1} ~~~~~~.
\end{eqnarray}
Hence, the density can be written in terms of the volume as :
\begin{eqnarray}
\rho & = & \left(\frac{B}{A + 1 - c_D \xi_0} + \frac{C}{a^{(D-1) (\alpha +1) (A + 1 - c_D \xi_0)}} \right)^{\frac{1}{\alpha +1}}  ~~~~~.
\end{eqnarray}
Interestingly, equation (24) reduces to the non-viscous MCG when $\xi_0$ is set to zero. In Fig.3, the variation of the energy density with respect to the scale factor has been plotted for different values of the viscosity parameter $\xi_0 =0$ (black color), $0.05$ (red color), $0.1$ (blue color), respectively. \\

\begin{figure}[hbtp]
\centering
\epsfxsize=6cm
\centerline{\epsfbox{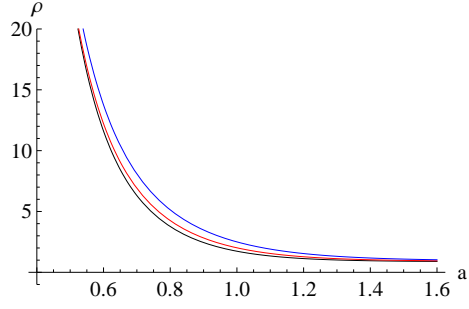}}
\caption{Plot of the density $\rho$ versus the scale factor $a$ for $\nu = \frac{1}{2}$ .}
\end{figure} 

The explicit form of the the time $t$ in terms of the energy density $\rho$ is given by: 
\begin{eqnarray}
c_D ~t + C  & = & \frac{2 \rho^{\alpha+\frac{1}{2}}}{B (2 \alpha +1)}~ 
 {}_2F_1 \!\left[ \begin{array}{cc}
1 ,1- \frac{1}{2 (\alpha +1)} \\
~~~2- \frac{1}{2 (\alpha +1)} \end{array}  ;
\frac{A+1 - c_D ~\xi_0}{B} \rho^{\alpha+1} \right] 
\end{eqnarray}
where $C$ is an integration constant. \\

\section{Statefinder Parameters in D Dimensions}
The expansion dynamics of the universe is probed through the derivatives of the scale factor with respect to the cosmic time. Indeed $\dot{a} >0$ means that 
the universe is undergoing an expansion and $\ddot{a} >0$ means that the universe is experiencing an accelerated expansion. \\

Geometrical parameters such as the Hubble parameter $H$ depending on the first derivative of the scale factor and the deceleration parameter $q$, 
\begin{eqnarray}
q & = & - \frac{\ddot{a}}{a H^2} 
\end{eqnarray}
depending on the second derivative of the scale factor are two elegant choices to describe the expansion state of the universe. The cosmic acceleration indicates 
that $q$ should be less than zero. \\

From equations (9) and (11), the acceleration parameter $q$ can be explicitly written in $D$ dimensions as : 
\begin{eqnarray}
q & = & \frac{D-3}{2} + \frac{D-1}{2} ~\frac{p_{eff}}{\rho} ~~~~.
\end{eqnarray} 
However both the Hubble and the deceleration parameters are not enough to characterize cosmological models uniquely because a quite number of models may just have 
the same current values of $H$ and $q$. \\

In order to be able to distinguish between various cosmological dark energy models, a robust diagnostic of dark energy based on higher derivatives of the scale 
factor has been proposed.  The statefinder pair $\{r,s \}$, which is a geometrical diagnostic, defines two new cosmological parameters in addition to $H$ and 
$q$. \\

In $D$ dimensions, the $r$ parameter will be defined as follws : 
\begin{eqnarray}
r & = & \frac{\dddot{a}}{a H^3} - (D-4) (q + 1) 
\end{eqnarray}
which by using equation (9) and (11) can be expressed in $D$ dimensions as :
\begin{eqnarray}
r & = & 1 + \frac{(D-1)^2}{2} \left(1 + \frac{p_{eff}}{\rho} \right) ~\frac{\partial{p_{eff}}}{\partial{\rho}} ~~~. 
\end{eqnarray} 

The latter equation suggests that the $s$ parameter has to be defined in $D$ dimensions as follows : 
\begin{eqnarray}
s & = & \frac{r - 1}{(D-1) ~(q - \frac{D-3}{2})} ~~~.
\end{eqnarray}
Using equations (27) and (29), the $s$ parameter becomes : 
\begin{eqnarray}
s & = & \left( 1 + \frac{\rho}{p_{eff}} \right) ~\frac{\partial{p_{eff}}}{\partial{\rho}} ~~~~~~.
\end{eqnarray}
Furthermore by combining equations (29) and (31), the ratio between the effective pressure $p_{eff}$ and the density $\rho$ can be written as : 
\begin{eqnarray} 
\frac{p_{eff}}{\rho} & = & \frac{2 \left( r - 1 \right))}{(D-1)^2 ~s} ~~~~~~~.
\end{eqnarray}
It is easy to see that the newly defined $r$ and $s$ parameters reduce to the usual one introduced in \cite{sahni} when the dimension $D=4$. \\

For the MCG with bulk viscosity, the derivative of the effective pressure $p_{eff}$ with respect to the density $\rho$ can be expressed as : 
\begin{eqnarray}
\frac{\partial{p_{eff}}}{\partial{\rho}} & = & A ~(\alpha + 1) - \frac{\alpha ~p_{eff}}{\rho} + (\alpha + \nu + \frac{1}{2} ) ~\frac{\Pi}{\rho} 
\end{eqnarray}
where 
\begin{eqnarray}
\frac{\Pi}{\rho} & = & - c_D ~\xi_0 \rho^{\nu- \frac{1}{2}} ~~~~.
\end{eqnarray}

Using equations (29)-(33), the following relation between $r$ and $s$ are derived : 
\begin{eqnarray}
2 \left(r - 1 \right) s^2 + 2 \alpha s \left(r - 1 \right) + 
4 \alpha \frac{\left(r - 1 \right)^2}{(D-1)^2} & = &
s \left( A (\alpha +1) + (\alpha +\nu + \frac{1}{2}) ~\frac{\Pi}{\rho} \right) \left( (D-1)^2 s + 2(r - 1) \right) ~~~.
%2 (D-1)^2 \left(r - 1 - (D-4) (q+1) \right) s^2 + 2 (D-1)^2 \alpha s \left(r - 1 - (D-4) (q+1) \right) + 
%4 \alpha \left(r - 1 - (D-4) (q+1) \right)^2 & = & \nonumber \\
%(D-1)^2 s \left( A (\alpha +1) - (\alpha +\nu + \frac{1}{2}) ~\frac{\Pi}{\rho} \right) \left( (D-1)^2 s + 2(r - 1 - (D-4) (q+1)) \right)
\end{eqnarray}

%\begin{figure}[hbtp]
%\centering
%\epsfxsize=6cm
%\centerline{\epsfbox{q_versus_volume_nuhalf.eps}}
%\caption{Plot of the deceleration $q$ versus the volume $V$ for $\nu = \frac{1}{2}$ and $D=4$.}
%\end{figure}

%\begin{figure}[hbtp]
%\centering
%\epsfxsize=6cm
%\centerline{\epsfbox{r_versus_volume_nuhalf.eps}}
%\caption{Plot $r$ versus the volume $V$ for $\nu = \frac{1}{2}$ and $D=4$.}
%\end{figure}

%\begin{figure}[hbtp]
%\centering
%\epsfxsize=6cm
%\centerline{\epsfbox{s_versus_volume_nuhalf.eps}}
%\caption{Plot of $s$ versus the volume $V$ for $\nu = \frac{1}{2}$ and $D=4$.}
%\end{figure}

%\begin{figure}[hbtp]
%\centering
%\epsfxsize=6cm
%\centerline{\epsfbox{r_versus_s_nuhalf.eps}}
%\caption{Plot $r$ versus $s$ for $\nu = \frac{1}{2}$ and $D=4$.}
%\end{figure}
In the following, we turn to the cosmic parameters $q$, $r$ and $s$ and investigate the evolution trajectories of these parameters. Figures shown with caption 
Fig. 4 describe the evolution of cosmic parameters of the viscous MCG model where $\nu = \frac{1}{2}$ and $D=4$, and the viscosity parameter 
$\xi_0 = 0$ (black color), $0.01$ (red color), $0.03$ (blue color), respectively. The top left, top right and bottom left Figures correspond to the 
evolution of the deceleration $q$, $r$ and $s$ parameters with respect to the volume $V$. From these Figures, one can observe that for the MCG model with and 
without viscosity, the statefinder parameters $\{r,s \}$ tend to $\Lambda$CDM fixed point $\{1, 0 \}$ in the future. \\
Furthermore, the $r$ and $s$ parameters, plotted in the bottom right Figure, shows that there is one asymptote to $s$-axis corresponding to : 
\begin{eqnarray}
r & = & 1 + \frac{(D-1)^2}{2} (\alpha +1) (A - c_D ~\xi_0) ~~~.
\end{eqnarray}

\begin{figure}[hbtp]
\centering
  \begin{tabular}{@{}cc@{}}
  \includegraphics[width=.4\textwidth]{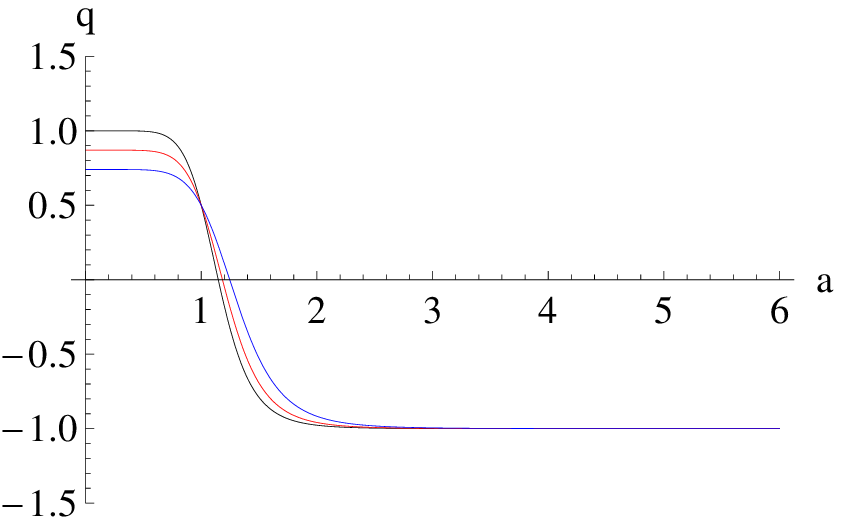} &
    \includegraphics[width=.4\textwidth]{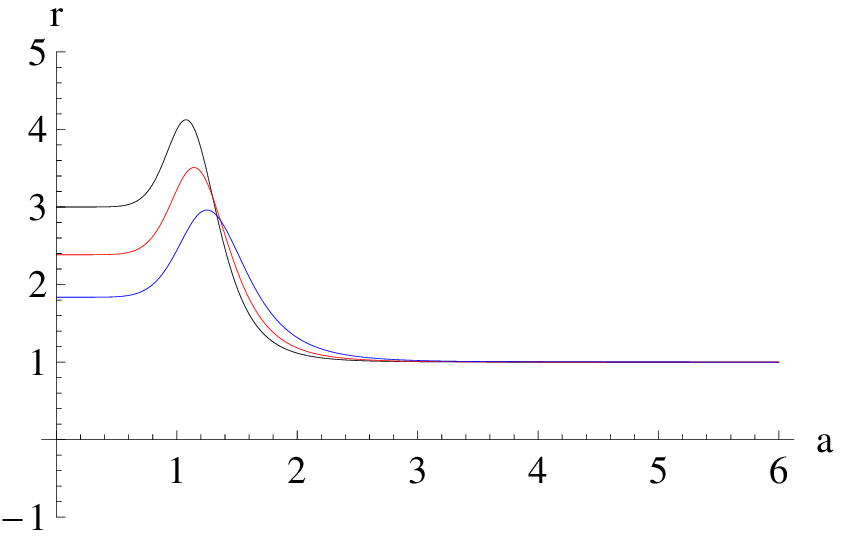} \\
    \includegraphics[width=.4\textwidth]{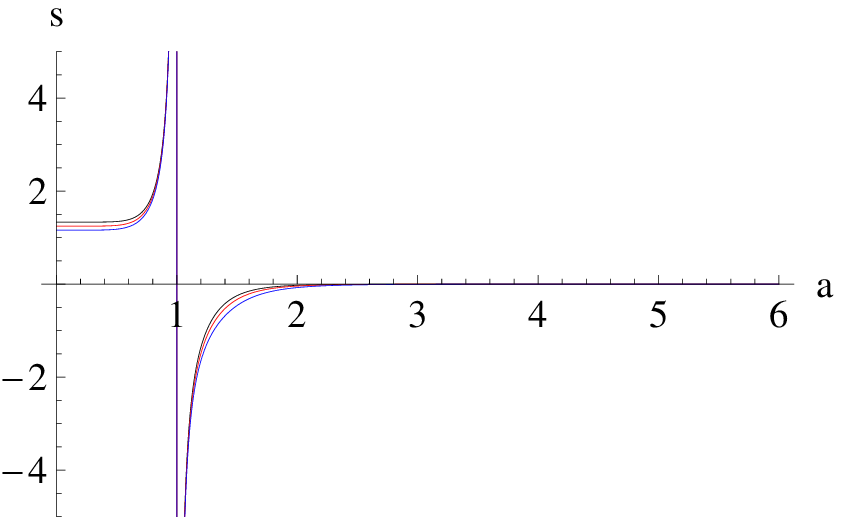} &
    \includegraphics[width=.4\textwidth]{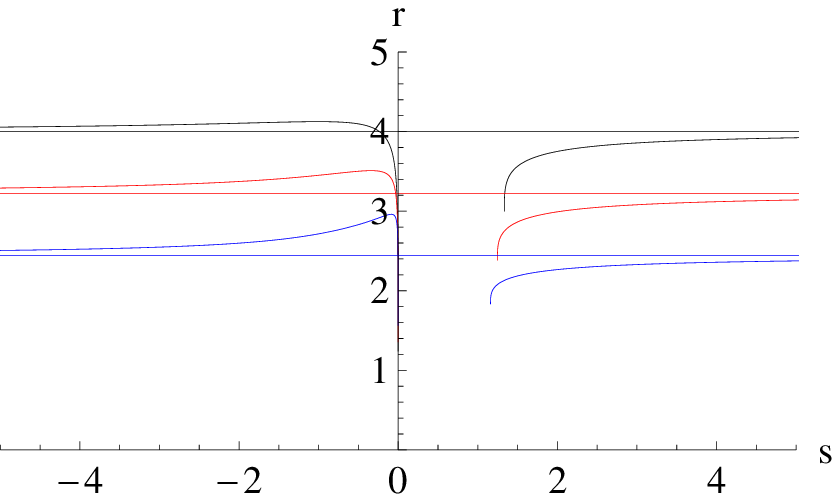} \\
\end{tabular}
\caption{({\em Top Left}) Plot of the deceleration $q$ versus the scale factor $a$. ({\em Top Right}) Plot of the $r$ parameter versus the scale factor $a$. 
({\em Bottom Left}) Plot of the $s$ parameter versus the scale factor $a$.
({\em Bottom Right}) $r$-$s$ diagram of the MCG model with and without viscosity.}
\end{figure}

%The statefinder pair $\{r, s \}$ of the viscous MCG for $D=5$ is displayed in Fig. 5 where the diagnostic pair $\{r, s \}$ will approach the 
%$\Lambda$CDM fixed point $\{1, 0 \}$ in the future. Obviously, the evolving trajectory of the viscous MCG model for $D=5$ in the $r$-$s$ diagram is not 
%quite different from the case $D=4$. 
%\begin{figure}[hbtp]
%\centering
%\epsfxsize=6cm
%\centerline{\epsfbox{r_versus_s_nuhalfdim5.eps}}
%\caption{$r$-$s$ diagram of the MCG model with and without viscosity for $D=5$.}
%\end{figure}

\section{Conclusion}
In the present paper, we have considered the physical model of a bulk viscous MCG filled flat homogeneous and isotropic universe. We have derived and formulated the basic equations governing this model. 
We have obtained solutions by using both analytical and numerical techniques. 
We have defined the deceleration parameter and the statefinder $\{r, s \}$ pair in D dimensions.
We have investigated the evolution of the MCG model with and without viscosity in the flat FRW universe from the statefinder point of view. Trajectories of these various parameters have been displayed graphically by using appropriate 
values of the constants to understand their behavior.

%%%%%%%%%%%%%%%%%%%%%%%%%%%%%%%%%%%%%%%%%%%%%%%%%%%%%%%%%%%%%%%%%%%%%%%%%%%%%%%

\end{document}